\documentclass[galaxies,review,accept,pdftex,oneauthor]{Definitions/mdpi} 
\graphicspath{{Figures/}}
\firstpage{1} 
\pubvolume{1}
\issuenum{1}
\articlenumber{0}
\pubyear{2025}
\copyrightyear{2025}
\externaleditor{Michele Bellazzini}
\datereceived{ } 
\daterevised{ } 
\dateaccepted{ } 
\datepublished{ } 
\hreflink{https://doi.org/} 

\usepackage{graphicx} 

\Title{Red and Yellow Hypergiants}
\TitleCitation{Red and Yellow Hypergiants}

\Author{{Terry Jones} \orcidA{}}

\AuthorNames{Terry Jones}

\AuthorCitation{Jones, T.}

\address[1]{%
{Minnesota Institute for Astrophysics, University of Minnesota, 116 Church St. SE, Minneapolis, MN 55455}; tjj@astro.umn.edu}  

\abstract{The red and yellow hypergiants are a rare and important phase in the evolution of the most massive stars that can reach the cool part of the HR Diagram. The hypergiant phase is commonly characterized by high, often episodic mass-loss rates and significant changes in spectral type, probably due to the formation of a pseudo photopsphere during a high mass-loss episode. Many of the yellow hypergiants are the immediate successors to the most luminous red supergiants, and often show evidence in their dusty, circumstellar envelopes from past red supergiant activity. In this paper we review the yellow and red hypergiants with an emphasis on how they differ from more normal red supergiants.}

\keyword{{massive stars; stellar evolution; stellar mass-loss}}  

\begin{document}


\section{Introduction} \label{sec:intro}

The term hypergiant has no official definition. In 1942 Keenan listed RW Cep as luminosity class Ia-0 based on specific line ratios in its spectrum, and suggested it was among the most luminous stars known \cite{keen42}. Feast and Thackery \cite{FT56} noticed that several stars in the Large Magellanic Cloud (which has a known distance) with spectral types from late F to middle G had absolute visual magnitudes of M$_V$$\approx -9$. These `super-supergiants' with M$_{Bol}< -9$ were given the name `hypergiant' by de Jager \cite{deJa98} and are commonly designated as luminosity class Ia0$^+$ or even I 0 in an extended MK classification system. Keenan \cite{keen71} originally suggested using the term for the intermediate temperature stars (yellow) stars that are both brighter than M$_V < -8.5$ to $-$9, and show a broad component of H$\alpha$ in emission, indicating a high mass loss rate. 


In this paper we review our knowledge of the most luminous stars cooler than late A in spectral type, redward from T$_{\rm{eff}}  \sim 8000$~K  along the Humphreys-Davidson limit in the HR diagram \citep{HD79}.  In this context we will apply the concept of hypergiant to stars more luminous than Log(M$_{bol}$)$= 5.4$, with T$_{\rm{eff}}<8000$~K, and evidence for significant mass loss, often episodic in character. This group includes both red and yellow hypergiants.


The most luminous stars between {$8000~ \rm{K} > \rm{T}_{eff} > 5000~\rm{K}$} (the yellow hypergiants) are very rare, due both to their high initial mass and rapid evolution in this portion of the HR diagram \cite{deJa98}. We include these stars in our discussion of the most luminous red supergiants because the most luminous RSGs are likely the immediate progenitors of the yellow hypergiants. Additionally, the yellow hypergiant phase is so short \citep{hump78, HD79}, that many retain evidence of very high mass loss from the immediately preceding red supergiant phase (e.g., \citep[][]{Shen16}). Confirming theoretical evolutionary time scales for massive stars by comparing blue and red supergiant numbers still suffers from incompleteness of the star counts in different temperature or spectral ranges. In his survey of yellow and red supergiants in M31 and M33, Gordon et al. \cite{gord16} concluded that 20--30\% of the YSGs were probable post-RSGs, consistent with the approximately equal times spent in each direction on the HRD according to the models.   


Study of these evolved massive stars is important. They return a significant amount of mass to the interstellar medium, are the progenitors of many types of core collapse supernovae, and create the immediate environment in which those supernovae explode. Although mass loss winds are ubiquitous in late type, luminous stars, the warm and cool hypergiants are distinguished by an observational record of large mass loss events in addition to the more normal mass loss processes. The underlying physics behind these energetic, episodic mass loss events is not understood but surface activity may be important. 

\section{General Properties}


\subsection{Observational Characteristics} 

\begin{figure}[H]
    \includegraphics[width=.9\columnwidth]{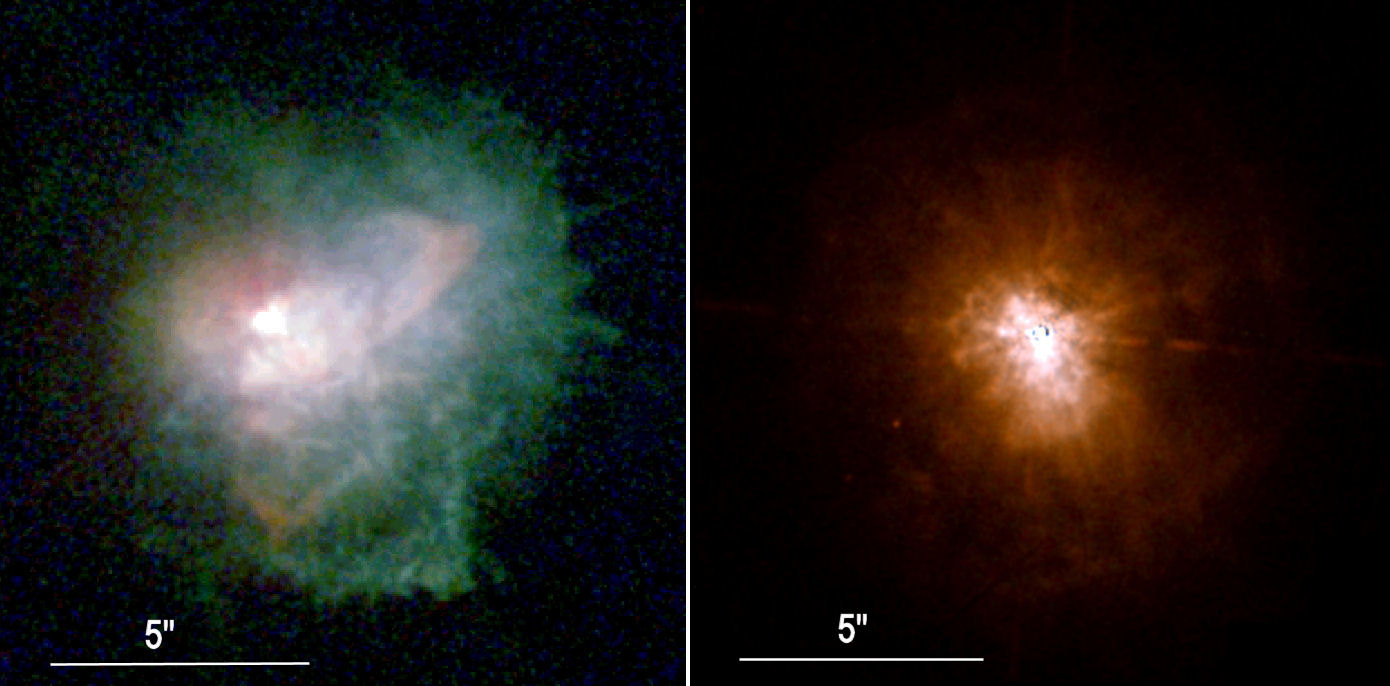}
    \caption{\label{fig:VY} \textbf{Left}: HST multiwavelength color image of VY CMa at optical wavelengths \cite{smit01}. \textbf{Right}:~HST image of IRC +10420 at optical wavelengths \cite{hump97}.}
\end{figure}

\begin{figure}[H]
    \includegraphics[width=.9\columnwidth]{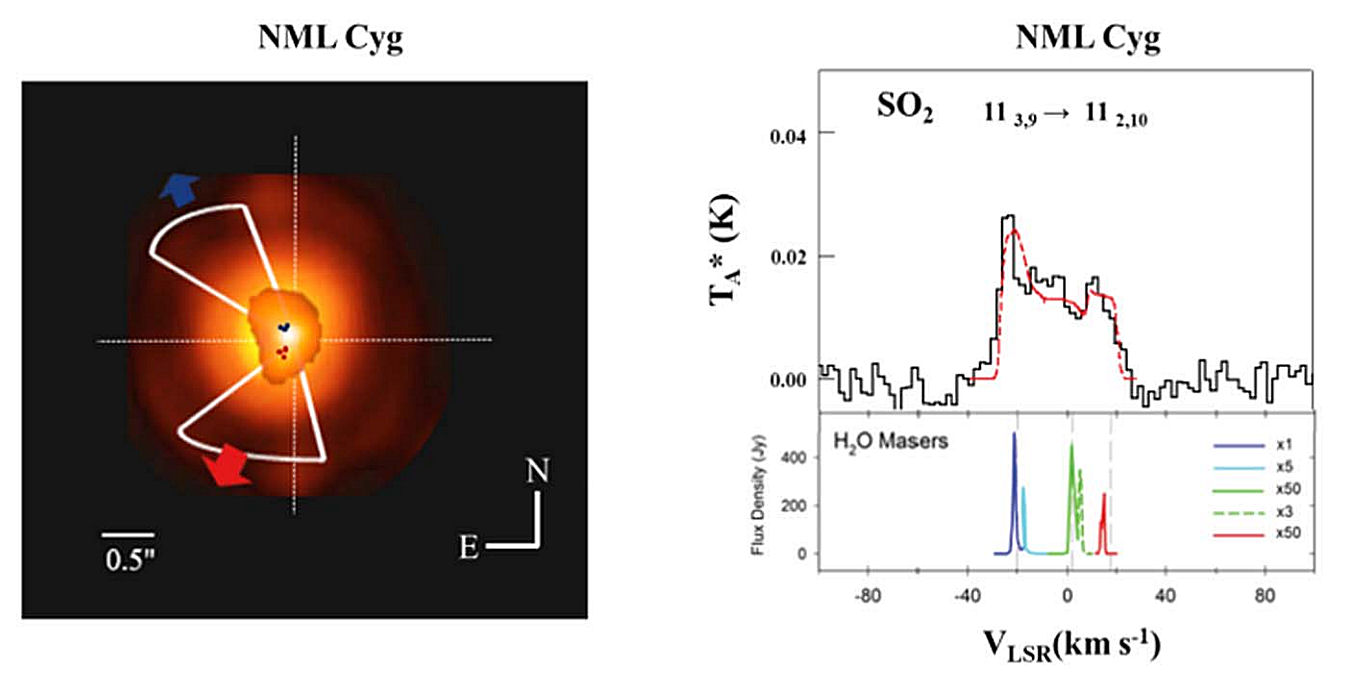}
    \caption{\label{fig:NML} \textbf{Left}: \{Proposed geometry of the collimated molecular outflows in NML Cyg, projected onto the plane of the sky, superimposed over the $11~\upmu\text{m}$ and HST images \citep{schu09}. White indicates the molecular outflow with the blue arrow designating blueshifted emission and the red arrow indicating redshifted emission. \textbf{Right}: Spectrum of SO$_2$. The water maser emission spectrum is shown below with the blueshifted (blue), centeral (green), and redshifted (red) components. Adapted from \cite{sing21}.} 
\end{figure}

The red hypergiants, like many RSGs, experience mass loss, but at rates that exceed what is common for those stars in the Milky Way and the Magellanic Clouds. Their spectral energy distributions (SEDs) show excess emission often from  2 $\upmu$m to more than 20 $\upmu$m indicative of extended circumstellar ejecta, which is also apparent in their optical images from HST and infrared imaging with SOFIA  \citep{Shen16, gord19}. The best known are VY CMa with its complex circumstellar ejecta \cite{smit01}  exhibiting multiple outflows \citep{hump05, hump07} and IRC +10420 showing arcs and complex structure structure \cite{hump97, tiff10} (Figure \ref{fig:VY}). Images of the red hypergiant  NML Cyg is altered by the winds from nearby hot stars \citep{schu09}, with two massive outflows seen in the mm-wave spectra and visible in its mid-infrared images (Figure \ref{fig:NML}).

\unskip

\begin{figure}[H]
    \includegraphics[width=\columnwidth]{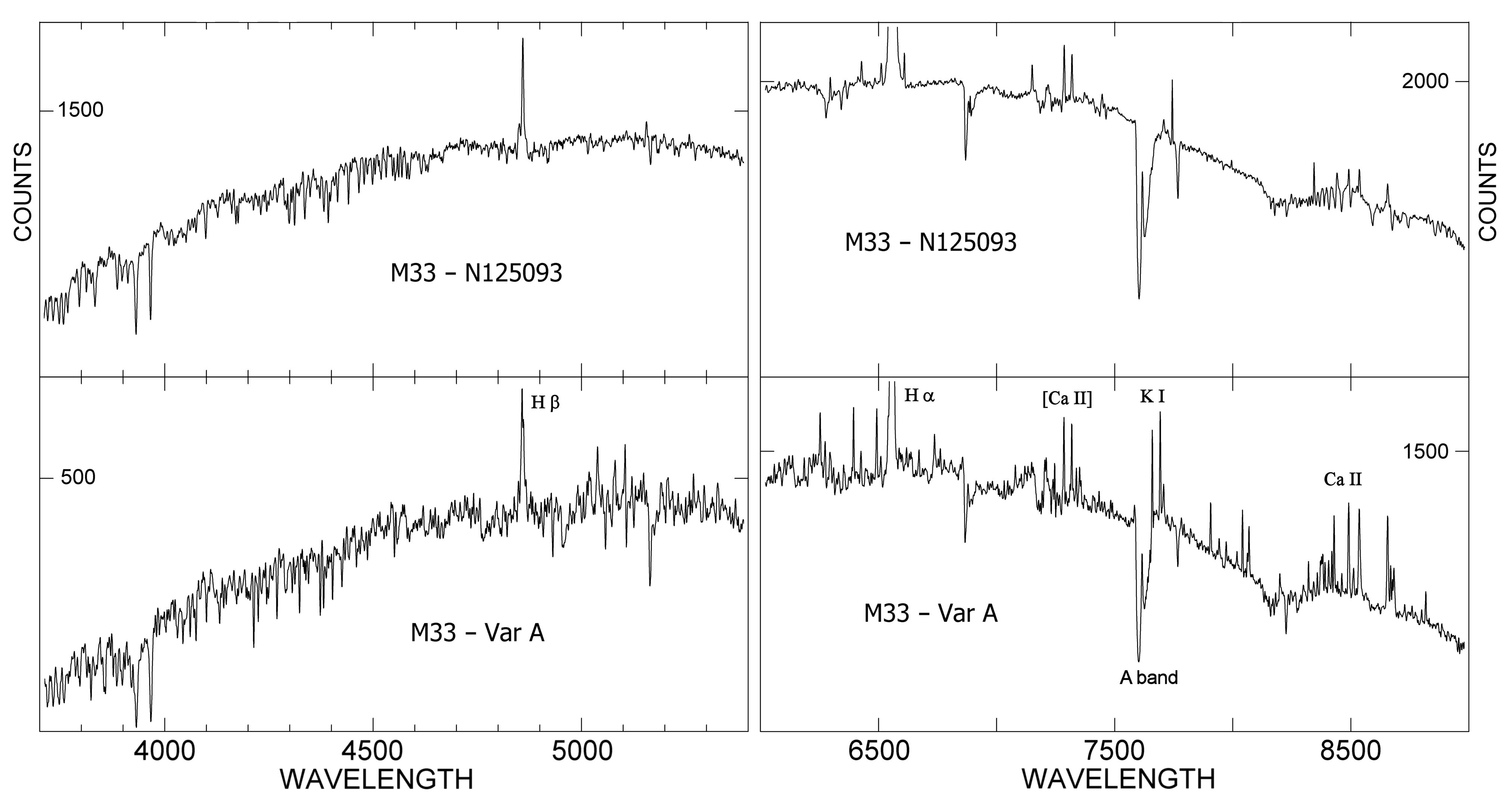}
    \caption{\label{fig:M33Spec} Spectra of warm hypergiants in M33, including Var A \cite{hump13}. Several spectral features that help identify hypergiants are shown.}
\end{figure}

The warmer hypergiants share these same features with infrared circumstellar emission revealed in their SEDs, and images showing numerous arcs, knots, and shells. IRC +10420, often called the prototype of the yellow hypergiants (Figure \ref{fig:VY}), and the “fried egg nebula” (IRAS 17163-3907) with multiple shells \citep{oudm22} are examples.  The spectrum of IRC +10420 is dominated by very strong H emission lines and emission in the Ca II infrared triplet with strong P Cygni  profiles and [Ca II] from  its extended ejecta \citep{oudm98, hump02}.  These characteristic emission lines are observed in the spectra of several yellow hypergiants in M31 and M33 (Figure \ref{fig:M33Spec}) which also have strong infrared emission from their dusty ejecta~\cite{hump02}.

\begin{figure}[H]
    \includegraphics[width=.85\columnwidth]{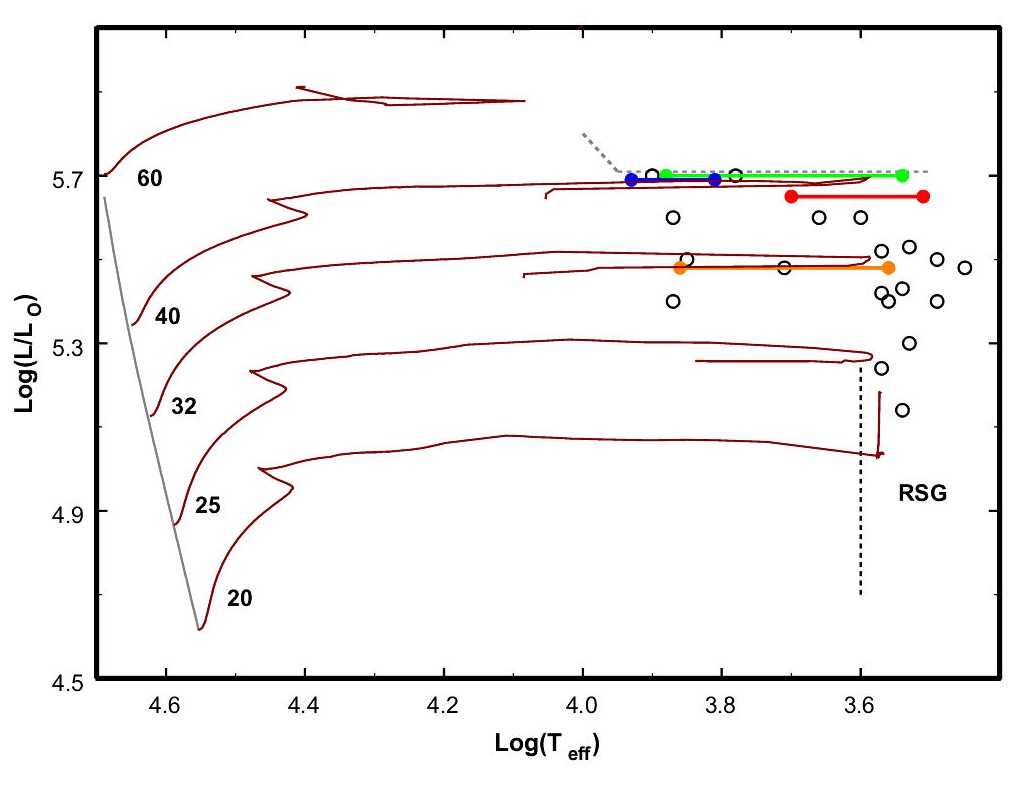}
    \caption{\label{fig:HR} HR Diagram with evolutionary tracks for 20 to 60 M$_\odot$ main sequence progenitor masses with Solar metalicity and no rotation \citep{ekst12}. For clarity, the tracks have been truncated after the end of the first `blue loop' back to hotter temperatures. The grey dashed line indicates the Humphreys Davidson Limit \citep{HD79}, the black open circles are from Table \ref{tab1} and the colored lines are the four stars discussed in Section~\ref{sec3}. Blue is IRC +10420, green is Var A, orange is $\rho$ Cas, and red is WOH G64.}  
\end{figure}

Table \ref{tab1} contains a selected list of red and yellow hypergiants. They are shown on an HR Diagram in Figure \ref{fig:HR}. This list is not intended to be exhaustive, but representative of the class. Note that not all of these stars have a secure hypergiant status. For example, if MY Cep is at the distance of NGC 7491 (which has a main sequence turnoff at $\sim$B1), based on its SED it must have a luminosity of L$_\odot \sim 3\times 10^5$ \cite{hump20}, suggestive of a significantly more massive 30M$_\odot$ star. The discrepancy between main sequence ages and RSG ages in clusters is discussed at the end of Section~\ref{sec2.2}.

\begin{table}[H]
\caption{Selected Hypergiants and Related Stars.\label{tab1}}
\begin{tabularx}{\textwidth}{LCCC}
\toprule
\textbf{Star}  &  \textbf{SpTy}  &  \boldmath{$log \rm{T}_{eff}$}  &  \boldmath{$log(L/L_\odot)$} \\
\midrule
Milky Way & & & \\
\midrule
AH Sco  &  M4-5 Ia-Iab  &  3.57  &  5.52 \\
HR 5171A & K0 0-Ia  & 3.70  &  5.70 \\
IRAS 17163-3907 & A3\-A6Ia & 3.90 & 5.70 \\  

IRC +10420 & F8Ia$^+$-A2I & 3.93-3.81 & 5.70 \\
KW Sgr  &  M0I-M4Ia  &  3.57  &  5.24 \\
$\mu$ Cep & M2Ia$^+$ & 3.57 & 5.42 \\
MY Cep & M7-M7.5  & 3.45 & 5.48 \\
NML Cyg & M6Ia & 3.49 & 5.50 \\
$\rho$ Cas & F8pIa-K0pIa & 3.86-3.56 & 5.48 \\
RSG C1 1--13 & M3/K2  & 3.71 & 5.48 \\
RSG C2 2--49 & K4 & 3.60 & 5.60 \\
RW Cep & K0-K2 & 3.71 & 5.48 \\
S Per & M4Ia & 3.53 & 5.30 \\
UY Sct &  M2-M4Ia-Iab &  3.53  &  5.53  \\
V602 Car  &  M3 Ia-Iab  &  3.54  &  5.14 \\
VX Sgr & M4Ia-(M8) & 3.49 & 5.40 \\
VY CMa & M4Ia$^+$ &3.54 & 5.43 \\
\midrule
LMC & & & \\
\midrule
HD 269953 & G0 & 3.78 & 5.70 \\
MG 73-59 & K0I & 3.66 & 5.60 \\
Sk-69-148 & K0I & 3.66 & 5.60 \\
WOH G64 & M-G: & 3.70-3.51 & 5.65 \\
\midrule
M31/33 & & & \\	
\midrule
M31-00444453 & F0Ia & 3.87 & 5.60 \\
M33 N093351 & F0Ia & 3.87 & 5.40 \\
M33 N125093 & F0-F2 & 3.85 & 5.50 \\
Var A & F0:-M4: & 3.88--3.54 & 5.70 \\ 
\bottomrule
\end{tabularx}
\end{table}

\unskip

\subsection{Evolution\label{sec2.2}}

The most luminous red and yellow hypergiants occupy a sparse, but important area in the HR Diagram \citep{HD79, gord19b}. As can be seen in Figure \ref{fig:HR}, they lie just below the empirical upper luminosity limit for stars with T$_{eff} \lesssim 8000$ \citep{HD79}. Evolutionary tracks \cite{ekst12} suggest they have main sequence progenitor masses between 25 and 40 M$_\odot$, and represent the most luminous stars possible in this temperature range. Stars more massive than 40 M$_{\odot}$ do not evolve to the red supergiant stage \citep{HD79}, most likely due to post main-sequence high mass loss driven by their proximity to the Eddington Limit \cite{davi20}. The high mass loss alters their interior structure, and they evolve back to higher temperatures without becoming red supergiants. The evolution of the RSGs is discussed in detail in the review by Ekstr{\"o}m and Gregory. 


The evolution of the hypergiants is governed by their initial mass, but mass loss can affect their evolution in the red and yellow hypergiant phases. Mass loss in RSGs is well known and well documented. Due to the lower surface gravity at these luminosities ($g \sim 0$), the outer circumstellar gas is only weakly bound to the star, a condition ripe for enhanced mass loss. Mass loss in red supergiants is covered in the review by Van Loon, and range from rather modest rates of $10^{-6} \rm{M}_{\odot}\rm{yr}^{-1}$ to as high as $10^{-4} \rm{M}_{\odot}\rm{yr}^{-1}$ in some of the yellow and red hypergiants (e.g., \citep[][]{hump20}).  


Initially, the yellow hypergiants, like HR 5171A, HR 8752 and $\rho$ Cas, known for their spectroscopic and photometric variability were assumed to be unstable because they were just below the upper luminosity limit.  Jones et al. \cite{jone93} presented strong evidence that the yellow hypergiant IRC+10420 is a post- red supergiant by analogy to the evolution of lower mass AGB stars, based on its high mass loss and extensive circumstellar ejecta. de Jager \cite{deJa98} has suggested that all yellow hypergiants are post-RSGs and in their evolution to warmer temperatures, they enter a period of dynamical instability (T$_{eff}$  $\sim$ 6000--9000 K) with enhanced mass loss. The evolutionary tracks used in Figure \ref{fig:HR} are for Solar metalicity, non-rotating stars. They show post-RSG blueward evolution, often called a "blue loop", when the He core mass exceeds a certain fraction of the star's mass. This is usually about two-thirds for RSGs with initial mass more than 25 M$_{\odot}$. The evolution  and interior structure of the RSGs is reviewed in the contribution to this issue from Ekstr{\"o}m \& Georgy.


IRC +10420 \citep{tiff10, Shen16}, has apparently made the transition between red supergiant to yellow hypergiant on the $\sim$1000 yr timescales expected. Other prominent examples are Var A in M33 and WOH G64 in the LMC. Var A was an F-type yellow hypergiant that transitioned to an M star, then after several decades returned to its prior warmer temperature  due to the dissipation of an optically thick wind from the previous high mass loss episode \citep{hump06}. Recent spectroscopic observations of WOH G64, formerly a very cool hypergiant in the LMC, show that the star has dramatically transitioned to a yellow hypergiant on a short timescale \cite{munoz}.


In a survey of Type II-P supernova progenitors, \citet{smartt} found evidence for an upper mass limit of only 16--17 M$_\odot$ for the RSG progenitors, leaving the more massive red supergiants, in particular the hypergiants, unaccounted for. It is possible that the more massive red supergiants have a different terminal state than Type II-P SN perhaps on a blue loop in the HRD, as a yellow hypergiant or blue supergiant followed by a terminal explosion, or direct collapse to a black hole. Several authors have questioned Smartt's conclusions arguing that incorrect determination of the star's SED due to lack of infrared data would lead to underestimates of the star's luminosity and therefore the mass estimate~\cite{beas16, beasor}. Smartt also considered this possibility, and other authors \citep{walm12, kilp18} have shown that it did not remove the discrepancy for stars above 20 M$_{\odot}$, now known as the red supergiant problem.  


Since massive stars are often in binaries, the effects of a binary companion could influence the evolution of the system just before, and during the hypergiant phase. Indeed, Beasor et al. \cite{beas19} find a systematic offset between ages determined using the main sequence stars in a cluster and using the evolved RSGs, suggesting binary mergers as one possible explanation. There are no measured binary motions in the known hypergiants \cite{deJa98}, but one system, HR 5171 \cite{ches14} has been imaged with sufficient spatial resolution to discover a hot, but low mass, companion. Chesneau et al. \cite{ches14} discuss the possible effects of even a low mass companion on the mass-loss process in the very extended atmosphere of the cool, and very luminous primary. They suggest behavior seen in HR 5171 is similar to other hypergiants, but clear evidence for binarity is still lacking. Whether or not a common envelope merger can explain the peculiarities of the hypergiants such as extreme, episodic mass-loss, is unknown. 

\subsection{High Spatial Resolution Observations}

With the advent of space based imaging, ground based optical and near IR adaptive optics and interferometry, and mm wave interferometry, it is now possible to explore the circumstellar ejecta, the surface, and the region in the mass loss wind near the surface associated with these stars. Depending on the facility, imaging that spans spatial scales from the inner mass-loss wind down to features on the surface of the actual star can be obtained. By covering a range of wavelengths in the visual, near and mid-infrared, mm waves and microwaves, surface phenomena, emission lines, dust emission, and maser emission are observable. The near circumstellar environments of RSGs is discussed in the review by Wittkowski. Here we concentrate on the hypergiants.


With the launch of HST, routine sub-arcsec imaging from the UV to the near-IR of the dust shells surrounding evolved stars became possible. Scattered light from the dust in the circumstellar environment of VY CMa \cite{smit01} and IRC+10420 \cite{hump97} revealed a very complicated mass distribution with knots and extended arcs  of emission, far from the spherical wind normally used in modeling the SEDs of these stars. A dust shell around IRAS 17163-3907 (Fried Egg Nebula) was not detected in HST images \cite{siod08}, but was later found using VLT/GRAVITY observations in combination with other data \cite{koum20}, and shows evidence for multiple shell ejections on 100 yr timescales.


Adaptive optics has greatly enhanced the imaging capability of ground based observations, particularly in the near and mid infrared. For example, Schuster et al.~\cite{schu09} imaged NML Cyg at mid-infrared wavelengths using the MIRAC3/BLINC camera \cite{hoff93} on the 6.5 m MMT telescope in combination other observations and found hot dust emission extending to the north west, likely heated by radiation from Cyg OB2. The same instrument was used by Shenoy et al. \cite{Shen16} to image the extended emission in IRC+10420 and other hypergiants. In light of recent variability seen in RW Cep, Jones et al. \cite{jone23} reanalyzed previous MIRAC observations and found the star had undergone a period of enhanced mass-loss about 100~yrs ago. AO imaging at near-infrared wavelengths has the advantage that circumstellar extinction is lower than in the visual, allowing deeper penetration into the dust shell, but without the difficulties of the high thermal background at MIR wavelengths. At optical wavelengths the image of VY CMa is dominated by scattered light from optically thin and thick lines of sight, but from 2.2--5~$\upmu$m most of this dust is optically thin, and only the  SW clump stands out~\cite{shen13}. This work was followed up with imaging at $10~\upmu$m by Gordon et~al.~\cite{gord19} who found that the SW clump was also optically thick in thermal emission.


In various configurations, the VLT has been able to make imaging and spectroscopic observations of hypergiants at high spatial resolution (e.g., see \cite{witt21, ohna24}).  Imaging polarimetry of VY CMa has found large dust grains were required to explain the scattered light~\cite{scic15}. Observations using he VLT/AMBER system have shown the presence of molecular layers of water vapor and CO in the star's extended atmosphere with an asymmetric morphology~\cite{witt12}.  Further observations of hyperegiants with the VLT include HR 5171 and IRC+10420~\cite{ches14, koum22}. 


Various techniques such as occultations and interferometry have been used to measure the diameters of cool luminous stars \cite{dyck95, klop15}, and  show that they did not have circularly symmetric limb darkening \cite{ragl06}. More recent results from CHARA \cite{anug24, anug24b, norr21} have been able to image the surface of AZ Cyg, RW Cep and $\rho$ Cas and have found very large and variable asymmetries, presumably the tops of large convection cells.  


ALMA imaging of VY CMa \cite{kami19, beck15} indicate a significant fraction of TiO$_2$ remains in the gas phase outside the dust-formation zone and suggest that this species might play only a minor role in the dust-condensation process around extreme oxygen-rich evolved stars. Singh et al. \cite{sing23} combine ALMA interferometry with single dish data to reveal previously unseen outflows extending $4''$--$9''$ from the star. These recent ALMA results prove that the envelopes surrounding evolved stars are far from homogeneous, and that a variety of dynamical and chemical processes dictate the wind structure \cite{wall17, dinh09, decin16}. 


The mass-loss winds of hypergiants often produce molecular maser emission that is generally more complex than the smoother, spherically distributed maser emission from the AGB OH/IR stars. Richter et al. \cite{rich13} used VLBI to study the SiO maser emission from VY CMa at milliarcsec (AU) resolution of spots ranging from 0--60 milliarcsec from the star. They found an overall maser morphology that contains large-scale features that are persistent over multiple years. Richards et al. and Murakawa et al. \cite{anit98, mura03} were able to measure proper motions in water masers in VY CMa and VX Sgr using MERLIN, and found a general pattern of expansion. Observing IRC +10420, Dinh-V-Trung et al. \cite{dinh17} used the Karl G. Jansky VLA and found that in contrast to the dust and Hydrogen emission forming a bi-polar system, the dense $^{28}$SiO$(J=1-0)$ emitting clumps are distributed throughout a roughly spherical envelope. In this special issue, Ziurys and Richards review the complex molecular and maser emission observed in the red hypergiants. 


The distances to Galactic hypergiants can be difficult to measure. VLBI observations of maser spots in VY CMa have enabled Choi et al. and Zang et al. \cite{choi08, zang12} to use classic seasonal parallax to determine the distance to the star to high accuracy, confirming its very high luminosity. Similarly Chen et al. \cite{chen07} were able to place VX Sgr in the Sgr OB1 Association. For these two stars, at least, their very high luminosity is not in doubt.

\subsection{Extreme Mass Loss and Massive Outflows\label{sec2.4}}

The measured mass loss rates in red supergiants have recently been shown to have a prominent upward turn to higher rates with higher luminosity in the LMC, SMC and Milky Way \citep{wen24, yang23, hump20}. Many RSGs and yellow hypergiants have high measured mass loss rates sufficient to alter their interiors and the relative mass of the core, for evolution back to warmer temperatures before their terminal state.  Studying the nature and mechanism of the high rates of mass-loss in the hypergiants is important  in the final evolution of these stars, either as SNe or on a ``blue -loop'' to higher tmeperatures as blue supergiants, LBVs, and or WR stars. It also determines the type of supernova explosion they will undergo, their final end as either neutron stars or black holes, and the immediate environment within which the supernova event takes place. 


The topic of mass-loss in the red supergiants is well covered in the reviews by Van Loon and Wittkowski in this special issue. Here we are interested in some of the characteristics of mass-loss in  several of the hypergiants that is more extreme than other cool, luminous stars. In particular, we want to explore the ejection of dense, massive outflows of gas for example in VY CMa \citep{hump05}, sometimes exceeding $10^{-2} \rm{M}_\odot$ \citep{ogor15, hump24, jone23}, on time scales of years. The hypergiants are grouped together based on observational characteristics, and may have both a diverse set of causes as well as different mixtures of causes for their mass-loss properties. Unlike the case for red giants and AGB stars, these mechanisms are not likely to be metalicity dependent. For example, normal Mira variables in the LMC have a wind velocity (and hence mass-loss rate) about half the Milky Way \cite{wood92}, but the LMC hypergiant WOH-G64 has a very dense mass-loss wind, similar to M33 Var A and the Galactic hypergiants \cite{munoz}. 


Large mass-loss events that are more spherical in nature (e.g., $\rho$~Cas \cite{,Shen16}) have been interpreted as due to large instabilities in the pulsation characteristics of the yellow Hypergiants~\citep{glaz24}. Another, largely spherical, mechanism is the 'wave-heating' model of Quataert \& Shiode \cite{quat12}, where a small fraction of the energy in vigorous core convection is deposited into the outer layers of the star. Wu \& Fuller \cite{wu21} discuss the properties of wave energy transport in single star SN progenitors as a possible explanation for the presence of circumstellar material surrounding many core collapse supernovae. 


The circumstellar ejecta surrounding stars such as VY CMa, and IRC 10420  are much more complex, with localized regions visible in their images as discrete knots, arcs, and loops of dusty gas, rather than shells produced by a more spherical mechanism. The underlying physics behind these mass-loss events is unknown. Humphreys and Jones~\cite{hj22} present evidence that gaseous outflows are the dominant mass-loss mechanism for the most luminous RSGs (the red hypergiants) and an important contributor to a more typical RSG like Betelgeuse \cite{kerv11} (see the review on Betelgeuse by Dupree \& Montarges). They conclude that gaseous outflows, related to magnetic fields and surface activity and comparable to coronal mass ejections, are a major contributor to mass loss from RSGs and the missing component in discussions of the mass-loss mechanism. The presence of large-scale spots on the photospheres of many luminous cool stars such as AZ Cyg, RW Cep and Betelgeuse~\mbox{\citep{anug24, norr21, jadl24}} could be due to massive, AU scale convection cells and be linked to discrete mass-loss events (see \citep[][]{anug24b, lope22}).


A problem with this suggestion is the lack of a well understood energy source. For example, the ALMA clumps \cite{hump24} in the circumstellar ejecta of VY CMa contain a $minimum$ of 2--4$~\times 10^{-2}\rm{M}_\odot$. Using the measured expansion velocities ranging from 20 to 50 km-s$^{-1}$, the corresponding kinetic energy is $3\times 10^{44}$ ergs.  Jones et al. \cite{jone07} found that the arc-like features in the mass-loss wind of VY CMa were coherent structures, not chance alignments of dense regions along the line-of-sight. If this type of discrete mass ejection is analogous to coronal mass ejections (CMEs) on the Sun, which are usually attributed to the dissipation of magnetic field energy for 10--100 Gauss in a $\sim$$0.1\rm{R}_\odot$ volume \citep{emslie}, they would require kilo-Gauss field strengths on AU diameter scales.   


CMEs on the Sun are usually (but not exclusively) associated with relatively brief and intense X-ray emission \citep{emslie}. Montez et al. \cite{montez} observed VY CMa with the XMM-Newton X-ray satellite observatory, but were unable to detect X-Ray emission. They estimated a mean surface magnetic field of only a few miligauss. Betelgeuse, a normal red supergiant, underwent a dimming in 2020 that was found to be due to a mass-loss event that ejected material that formed dust and partially obscured the surface of the star \citep{drev24}. Although something similar to a CME could be responsible,  Kashyap et al. \cite{kash} found no evidence for X-Ray emission from the star at an earlier epoch. It is possible that these null results are simply due to the very sporadic nature of these events, and that these stars were X-Ray quiet at the time of observations. It is important to emphasize these X ray observations  were after the outflow events, many years for VY CMa and weeks for Betelgeuse.


Due to line broadening by macroturbulence, it is difficult to measure the surface magnetic field strength in convective stars such as red giants and RSGs using the Zeeman effect. Advances in this technique using multiple absorption lines, including blends, have resulted in the determination of the field strength in a handful of red giants and a few luminosity class Ib RSGs (see \citep[][]{plach} for a review). The typical surface magnetic field strengths range from 3--15 Gauss, which, since they represent averages across the entire surface of the star, do not preclude locally stronger field strengths. The polarization of molecular maser emission from well out into the circumstellar environment of red hypergiants such as VX Sgr and VY CMa can be interpreted as due to a magnetic field in the mass-loss wind (e.g.,~\citep[][]{vlem02, vlem05}). Extrapolation of the computed field strengths back to the photosphere assuming an $r^{-2}$ trend, suggests surface fields in the 100--500 Gauss range, less for an $r^{-1}$~trend.


Looking much closer to the star, Vlemmings et al. \cite{vlem17} detected polarized thermal dust emission from clump C, a dense volume of gas and dust $\sim$440 AU into the wind of VY CMa. Interpreting the polarized emission from the dust as due to grains aligned to the local magnetic field by the radiative torque process (see \citep[][]{ande15}), they derive a field strength \mbox{${\rm B}>13$ mG} in the clump. They also find the SiO maser emission closer in to the star to be circular polarized. If the polarization is due to the Zeeman effect, this would imply a magnetic field strength of 1--3 Gauss roughly 50--100 AU from the star. Extrapolating these results back to the star (R$_* = 6.5$ AU) suggests mean field strengths of \mbox{60--150 Gauss}. Different measurements of the magnetic field strength at different locations in the circumstellar environment using different techniques do not all extrapolate back to the photosphere with similar results. None-the-less, a mean surface field strength of a $\sim$100~Gauss seems possible. How a very slow rotating core in a RSG could generate such surface fields is a puzzle, but it has been suggested for Betelguese, for example, a local dynamo associated with giant convection cells can work \cite{auri10}. 


It is important to keep in mind that the magnetic energy available to power CMEs in the Sun is not well represented by a mean field strength, rather it is the excess “non-potential” magnetic energy, the energy above the minimum-energy potential (i.e., current-free) field to which the field can relax \cite{emslie}. This ``free'' magnetic energy derives from complicated field line geometries that may not contribute much to a measure of the mean field strength for the entire star. In other words, AU size regions on the surfaces of the hypergiants corresponding to large scale convection cells may contain considerable magnetic energy in strongly twisted magnetic field lines, even though measured global field strengths are `only' $\sim$10--100 Gauss. However, a stronger mean field strength should enhance the development of "non-potential" magnetic energy in the large scale convection cells seen in luminous, cool stars.

\section{Transits on the H-R Diagram and Post RSG Evolution\label{sec3}}

Three hypergiants are now found to have observed transitions in their apparent spectral types that include a red supergiant state---$\rho$ Cas, Var A in M33 and most recently WHO G64 \citep{munoz}.  The time scales are too short to be due to actual evolution from a red supergiant to a yellow hypergiant.  Instead, in $\rho$ Cas and Var A we actually observe the star transit from its normal yellow hypergiant state to a red supergiant with TiO bands then back to its warmer state.   These transitions are shown as dashed lines on the HRD in Figure~\ref{fig:HR}. 


Three of these episodes have been observed in $\rho$ Cas in the past century;  1946--1947~\citep{bide57}, 1985--11986~\citep{boya88} and 2000--2001~\citep{lobe03}.  These enhanced mass loss episodes produce a cool dense wind with TiO bands, accompanied by a dimming of 1 to 2 mag in the visual due to the shift from a yellow to red star.  The duration is relatively short, lasting only a few hundred days, and the mass loss rate briefly increases from $\approx$10$^{-4}$ to 10$^{-2}$ M$_{\odot}$ yr$^{-1}$.  The origin of these events is attributed to pulsation but the dynamics are not well understood in this part of the HRD.  Large convective regions have recently been reported on $\rho$ Cas \citep{anug24b} with an extended circumstellar envelope.  Its mass loss events may be driven by surface activity as suggested for other hypergiants in Section~\ref{sec2.4}.


Var A  in M33 was one of the original Hubble-Sandage variables, and was one of  the visually brightest stars in M33  as a yellow hypergiant until it faded four magnitudes in 1951--1953 due both to the shift in bolometric correction to a cooler star and to dust formation \cite{hump87, hump06}. A spectrum obtained in 1985 showed TiO bands, probably  due to a cool, dense extended wind. Together with its very red color and large infrared excess due to circumstellar dust (which still obscured the star), Var A looked like a RSG.   Beginning in about 2000, or before its B-V color had shifted from a red 1.7 mag to 1.0 mag and 0.8~mag in 2003, spectra from 2003 and 2004 showed that Var A had returned to its normal yellow hypergiant state with absorption lines consistent with an early F-type spectrum, plus strong H, Ca II, [Ca II], and K I emission lines. However, the star was sill visually faint. Thus it was still obscured by circumstellar dust along our line of sight.  Var A’s high mass loss episode lasted at least 35 yrs, possibly 45 yrs.  This is much longer than $\rho$ Cas’s short transitions, and implies a much more energetic outflow with higher mass to create the cool dense wind lasting at least 35 yrs. 


WOH-G64 has been recognized for some time as one of the most luminous RSGs in the LMC. It was recently observed in transition from an RSG spectrum to a yellow hypergiant~\citep{munoz} with corresponding changes in its light curve.  It was apparently in the red hypergiant state for at least 30 yrs, similar to what we observe for Var A. WHO-G64 had an M5 type spectrum in 2008 but in 2016 its spectrum had transitioned to a yellow hypergiant. Based on its light curve, the transition to a warmer temperature began earlier,  2014--2015 and may have been rather brief, on the order of a year. This may be seem rather short, but  assuming that the outer cool envelope is a wind, not a true photosphere, this transition could be similar to the change seen in Var A.  Humphreys et al. \citep{hump06}  showed that Var A’s transitions could easily occur on timescales of one year. We suggest that WHO-G64 is similar to Var A with a normal or quiescent state as a yellow hypergiant, and a probable post-RSG, that undergoes large mass-loss events producing a pseudo-photosphere with a cooler M type spectrum.  The available published optical photometry for WHO-G64 begins about 1993, and Elias et al.~\citep{elia86} report an M7.5 spectral type in 1986. Evidence for a warmer prior state  would require examination of the earlier plate material \cite{Westerlund}.  


We suspect that the yellow hypergiants showing short-term transits are very likely in a post-red supergiant state with their instabilities, mass loss, and circumstellar dust having characteristics shared with the red hypergiants. It isn't known what fraction of yellow supergiants may be post-RSGs. Gordon et al. \cite{gord16} found that 30 to 20\% of the known yellow supergiants in the nearby galaxies M31 and M33 had emission from stellar winds in their spectra and circumstellar dust which distinguished them from the YSGs with normal spectra and SEDS. They suggested that these stars were candidates for post-RSG evolution. Likewise, Humphreys et al. \cite{hump23} identified six of the most luminous yellow hypergiants in the LMC as probable post-RSGs, and Dorn-Wallenstein et al. \cite{dorn23, dorn20} has suggested that a small group of yellow hypergiants with short term variability in the LMC are post-RSGs.

\begin{figure}[H]
    \includegraphics[width=\columnwidth]{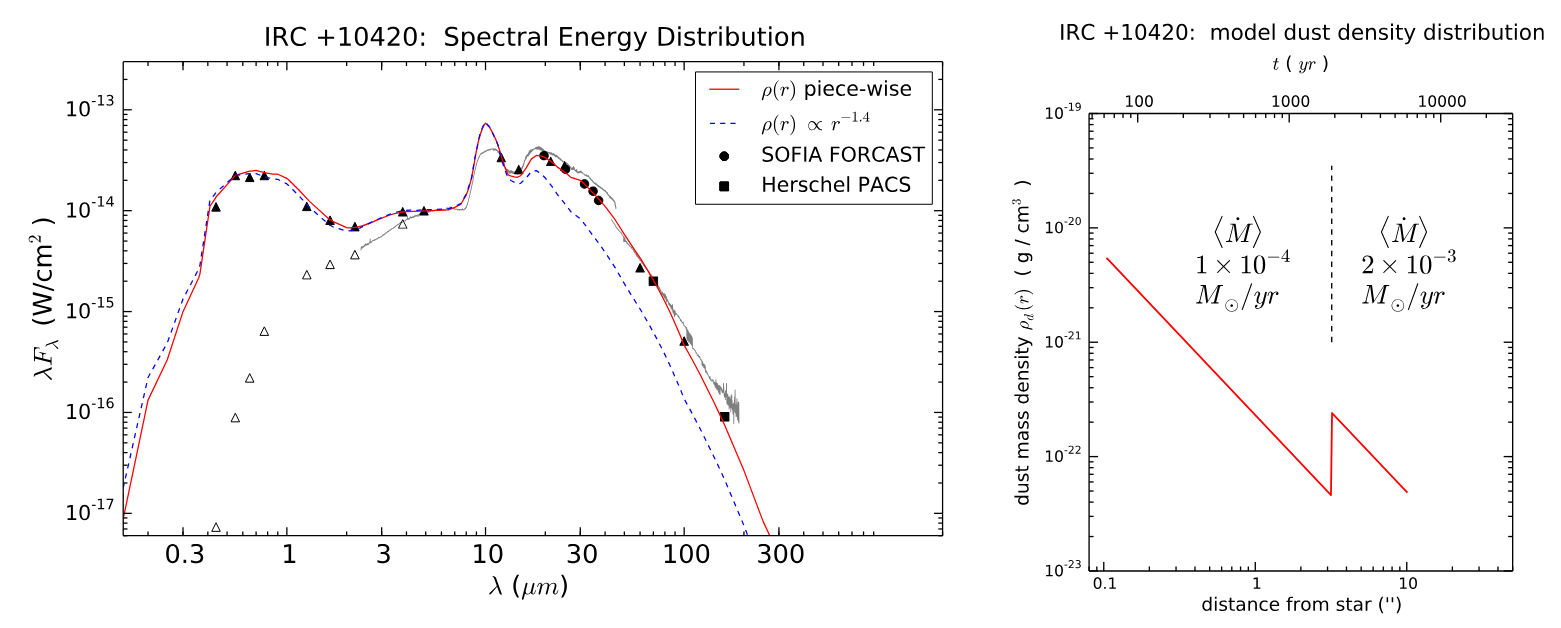}
    \caption{\label{fig:IRCSED}  The SED of IRC +10420 on the left and the mass-loss model used to fit the data. Adapted from \cite{Shen16}.}   
\end{figure}

Finally, one of the most studied yellow hypergiants is IRC +10240,  discussed in some detail in Gordon \& Humphreys \cite{gord19b}, is perhaps the best post-red supergiant candidate. Here we describe the characteristics of this star in the context of the hypergiants as in transition to a blue supergiant and likely core collapse supernovae. Using interstellar polarization, interstellar extinction and the radial velocity, Jones et al. \cite{jone93} were able to argue that the star was $\sim$5~kpc distant and a true hypergiant.  In 1964, IRC +10420 was listed as variable star SON 8102 in \cite{hoff64} and characterized as ``slowly changing'', but with no color or spectral type information. Its light curve shows a gradual one magnitude brightening over about 20 years beginning about 1920 \cite{gott78}. Spectra by Humphreys et al. \cite{hump73} indicated a spectral type F8 or G0 of high luminosity. Later, Oudmaijer et al. \cite{oudm98} found that the spectrum had changed to an early A spectral type in about 25 years most likely due to variations in its wind \cite{hump02}. Careful analysis of the SED by Shenoy et al. \cite{Shen16} (Figure \ref{fig:IRCSED}) found clear evidence for a past high mass loss phase, likely during the preceding RSG phase. 

\begin{figure}[H]
    \includegraphics[width=\columnwidth]{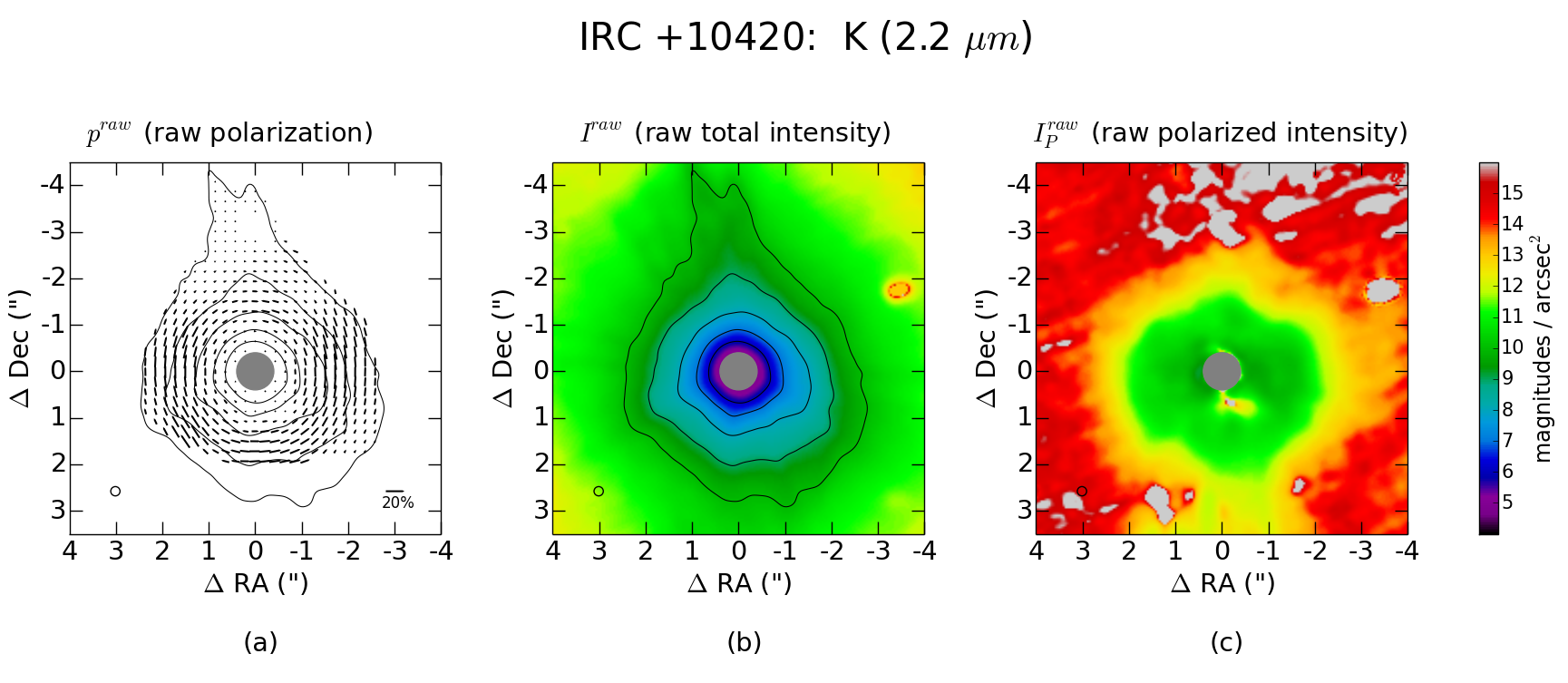}
    \caption{\label{fig:IRCPol} Imaging polarimetry of IRC +10420 at $2.2~\upmu$m. (\textbf{a}): Polarization half-vectors showing centro-symmetric scattering. (\textbf{b}): Total intensity (the center is saturated). (\textbf{c}): Polarized intensity showing the relatively flat surface brightness, face-on, excretion disk \cite{shen15}.} 
\end{figure}

Images of IRC +10420 show evidence for numerous, discrete outflow events \cite{oudm22, koum22, tiff10}, including compact arcs and knots with different space velocities and ejection history. Oudmaijer et al. \cite{oudm94} found the first evidence the star was embedded in a bipolar system with an asymmetric wind. This geometry has since been well established and determined to be pole-on \cite{davi07,tiff10, oudm22}. Using imaging polarimetry, observations by Shenoy et al. \cite{shen15} (Figure \ref{fig:IRCPol}) clearly reveal an optically thick face-on excretion disk. If the star continues to evolve to the blue, the current high mass-loss should decline, leaving an expanding ring left over from the thick excretion disk. More distant rings at the two openings of the bipolar cones, expanding at a greater rate, are likely.

\unskip

Interestingly, this geometry is just what we see in the material lit up by the explosion of SN 1987A \cite{mats24}. These rings were created about $2\times 10^4$ years ago \cite{crot91}, usually considered to have been during the RSG phase. Mechanisms for their formation is often postulated to have been caused by a rapidly rotating star (e.g., \cite{chit08}), a bipolar outflow from a binary system (e.g., \cite{akas15}), or a binary interaction (see Podsiadlowski for a review \cite{pods92}).  The yellow hypergiant HR 5171A may be in a contact binary \cite{ches14}, but the binarity of IRC +10420 is unknown. Whatever the mechanism for the formation of the bipolar geometry in the mass distribution around IRC +10420, it is entirely plausible that the rings seen in SN 1987A were produced during the yellow hypergiant phase of the progenitors evolution, not the RSG phase. 
The RSGs as supernova progenitors is discussed in detail in the review by Van~Dyk.   

\section{Summary}

The yellow and red hypergiants represent a diverse collection of stars with luminosities near the HD limit, complex mass-loss winds, and peculiar spectra compared to the less luminous, more normal supergiants in the same temperature range. Their location in the HR diagram strongly implies they are massive stars rapidly evolving between hotter and cooler temperatures. Some yellow hypergiants show clear evidence in their mass-loss history of a preceding red hypergiant phase, indicating they are currently evolving to the blue. The mass-loss winds of several hypergiants show very complex structure indicative of discrete, high mass ejections on time scales of $\sim$100 years. This type of mass-loss event is also found in more normal RSGs, such as Betelgeuse, but on a much reduced mass and energy scale. The origin of these discrete gaseous outflows is unknown, but the observation of large ($\sim$few AU) size spots on RSGs and some hypergiants suggests massive convection cells are involved, perhaps storing energy in twisted magnetic fields.

\vspace{6pt} 
\funding{This research received no external funding.} 

\conflictsofinterest{The authors declare no conflicts of interest.} 

\abbreviations{Abbreviations}{~The following abbreviations are used in this manuscript:\\

\noindent
\begin{tabular}{@{}ll}
HRD & Hertzsprung-Russell Diagram, H-R Diagram\\
RSG & red supergiant \\
SED & spectral energy distribution \\
VLBI & Very Long Baseline Interferometry \\
AU & Astronomical Unit \\
\end{tabular} 
}

\reftitle{References}

\begin{adjustwidth}{-\extralength}{0cm}

\PublishersNote{}
\end{adjustwidth}

\end{document}